\documentclass[aps,pra,twocolumn,superscriptaddress]{revtex4-2}

\usepackage[colorlinks=true,citecolor=blue]{hyperref}
\usepackage{graphicx}
\usepackage{amsmath}
\usepackage{amssymb}
\usepackage{bbold}
\usepackage{comment}
\usepackage{ulem} 

\DeclareGraphicsExtensions{.png,.jpg,.eps}
\usepackage{xcolor}

\begin{document}

\author{Wojciech Brzezicki}
\affiliation{International Research Centre MagTop, Institute of
  Physics, Polish Academy of Sciences, Aleja Lotnik\'ow 32/46, PL-02668 Warsaw, Poland}
\affiliation{Institute of Theoretical Physics, Jagiellonian University, ulica S. \L{}ojasiewicza 11, PL-30348 Krak\'ow, Poland}

\author{Timo Hyart}
\affiliation{Computational Physics Laboratory, Physics Unit, Faculty of Engineering and Natural Sciences, Tampere University, FI-33014 Tampere, Finland}

\title{Geometric and conventional contributions to  superfluid weight in the minimal models for superconducting copper-doped lead apatite}

   
\begin{abstract}
The density functional theory calculations and tight-binding models for the copper-doped lead apatite support flat bands, which could be susceptible to the emergence of high-temperature superconductivity. We develop theory for the geometric contribution of the superfluid weight arising from the momentum-space topology of the Bloch wave functions of these  flat bands, and we compare our results to the paradigmatic case of $s$-wave superconductivity on an isolated topological flat band. We show that, in contrast to the standard paradigm of flat-band superconductivity, there does not exist any lower bound for the superfluid weight in these models. Moreover, although the nontrivial quantum geometries of the normal state bands are the same when the superconductivity appears in the ferromagnetic and paramagnetic phases, the emerging superconducting phases have very different superfluid weights. In the case of superconductivity appearing on the spin-polarized bands the superfluid weight varies a lot as a function of model parameters. On the other hand, if the superconductivity emerges in the paramagnetic phase the superfluid weight is robustly large and it contains a significant geometric component.   
\end{abstract}

\maketitle


 \section{Introduction}

The reports of signatures of the high-temperature superconductivity in copper-doped lead apatite \cite{lee2023roomtemperature, lee2023superconductor} have stirred an active debate and ongoing attempts to replicate the results with so far only partial success  \cite{wu2023successful, hou2023observation}. Density functional theory (DFT) calculations   suggest that this material supports spin-polarized bands at the Fermi level \cite{griffin2023origin,si2023electronic, kurleto2023pbapatite, cabezasescares2023theoretical, jiang2023pb9cupo46oh2}, and because of their small bandwidth they are  called flat bands. It was proposed that the flat bands could support  correlated phases at high temperature \cite{griffin2023origin}, because the large density of states can lead to an exponential enhancement of the critical temperature \cite{Kopnin2011}. However, since the kinetic energy is quenched the flat bands are expected to be susceptible to various instabilities, which may lead to the appearance of magnetic, orbital or superconducting order \cite{Moon1995, jain_2007,  Annica2017, Ojajarvi2018, Hu2020, Lau21} as has been observed in quantum Hall systems \cite{Moon1995, jain_2007}, moir{\'e} superlattices~\cite{Cao2018, cao2018correlated, Yankowitz19, Lu2019, Stepanov20, Saito20, Serlin2020, Chen2020,Park21, Hao21, Cao-unconventional21} 
and other flat-band systems \cite{Zhou21, Zhou22}. One of the most exciting aspects of copper-doped lead apatite is that, in contrast to the flat-band systems studied experimentally so far, the flat bands in this system are three-dimensional, allowing the possibility of more stable correlated phases \cite{Lau21}. However, the understanding of the details of the competition of the order parameters as a function of external parameters and structural properties of the samples, as well as the experimental advances in the sample fabrication techniques to achieve robust control of the appearing order parameters, are major challenges for the future investigations. The guiding principle is that one should avoid being at half-filling of the flat bands if one wants to realize superconducting phases because the change of the filling away from the exact half-filling is more detrimental to the other types of order parameters than to the superconductivity \cite{Annica2017, Lau21}. In the spin polarized case the two spin-polarized flat bands are half filled on average (i.e. the total filling of the two bands is $1$) because the corresponding bands in the other spin sector are far below the Fermi level and fully occupied, but in the paramagnetic phase of this material the two flat bands in each spin sector are $3/4$ filled on average (i.e. the total filling of the four bands is $3$). Additionally, from the two flat bands the upper one has significantly larger bandwidth \cite{griffin2023origin,si2023electronic, kurleto2023pbapatite, cabezasescares2023theoretical, jiang2023pb9cupo46oh2}, and another important guiding principle is that the larger bandwidth also favours superconducting order parameter relative to the other order parameters. Thus, both guiding principles suggest that the paramagnetic phase is more favourable environment to support superconductivity, but this comes with the cost of lowering the overall magnitude of the critical temperature of the correlated phases, which is expected to reach the maximum at half-filling \cite{Annica2017, Lau21} and it decreases also with the increasing bandwidth \cite{Kopnin2011}. 

In order to understand the potential of the proposed phases of copper-doped lead apatite for high-temperature superconductivity, the key property which should be studied is the   
superfluid weight $D_s$, which captures the ability of a material to support supercurrent and the Meissner effect~\cite{Scalapino1992, Scalapino1993}. 
For conventional superconductors originating from the metallic state one has the well-known result
$D_s =e^2 n/m^*$,
where $n$ is the electronic density and $m^*$ the effective mass~\cite{Scalapino1992}. However, the experimental observations of superconductivity in twisted bilayer graphene~\cite{Cao2018, Yankowitz19, Lu2019, Stepanov20, Saito20} and other flat-band systems \cite{Park21, Hao21, Zhou21, Cao-unconventional21, Zhou22}, where the effective mass $m^*$ is large and one would expect $D_s$ to be small, demonstrate that there must exist also other contributions to the superfluid weight. Indeed, it has been theoretically found that, besides the band dispersion, also the quantum geometry of the Bloch wave functions contributes to the superfluid weight~\cite{Moon1995, Peotta2015, Liang2017, Hu2019,Fang2020,Julku2020,Hu2020, Rossi2021,Torma2021}. 
In particular, in the paradigmatic case of a well-isolated flat band supporting time-reversal invariant $s$-wave singlet superconductivity with spin-rotation symmetry around the $z$-axis 
the superfluid weight originates purely from the quantum geometry and there exist a lower bound determined by the integral of the absolute value of the Berry curvature \cite{Peotta2015, Liang2017}. This means that if this type of system supports spin-up and spin-down flat bands carrying opposite spin-Chern numbers there exists 
a lower bound for $D_s$ given by the value of the Chern number \cite{Peotta2015}. Similar lower bounds
can be obtained also when the bands are characterized by other topological invariants~\cite{Fang2020}, and significant geometric contributions have been found also in three dimensional systems \cite{Lau21} and disordered systems \cite{Lau22}. 

In this paper we study the geometric and conventional contributions to superfluid weight in the minimal models for superconducting copper-doped lead apatite 
\cite{Hirchmann2023, lee2023effective, tavakol2023minimal, zhou2023cusubstituted}. These models support nontrivial momentum space topologies \cite{Hirchmann2023,  zhou2023cusubstituted}, and therefore there could exist a significant geometric contribution to the superfluid weight. However, the studied models fall outside the standard paradigms of flat-band superconductivity because they can break all possible assumptions required for the existence of the universal lower bounds determined by topological invariants: (i) even though the bandwidths are small the bands are not perfectly  flat, (ii) the bands are not well-isolated and depending on the tight-binding parameters there can even exist topologically protected Weyl points which force the bands to be degenerate at certain momentum values \cite{Hirchmann2023,  zhou2023cusubstituted}, (iii) according to the DFT predictions the system does not obey time-reversal symmetry but instead the flat bands are expected to be spin-polarized \cite{griffin2023origin,si2023electronic, kurleto2023pbapatite, cabezasescares2023theoretical, jiang2023pb9cupo46oh2}, and (iv) the spin-polarized bands cannot support singlet superconductivity. We find that because of these reasons there does not exist any kind of lower bound for the superfluid weight in this kind of system. In particular, we show that the superfluid weight can even be negative in the proposed models, indicating instability of the superconducting state.   

We perform a comprehensive study of the behavior of the superfluid weight in the candidate models for superconducting copper-doped lead apatite by  considering both the interorbital-pairing  superconductivity in the spin-polarized phase \cite{tavakol2023minimal} and the $s$-wave singlet superconductivity in the paramagnetic phase. Although the paramagnetic phase is not supported by the DFT calculations there could exist competition between ferromagnetism and superconductivity so that the transition to the superconducting phase would be accompanied by disappearance of the spin polarization. Additionally, we also consider the effects of the spin-orbit coupling \cite{zhou2023cusubstituted} and mirror symmetry breaking terms \cite{Hirchmann2023} on the normal state, superconducting state and superfluid weight.  Interestingly, we find that  although the nontrivial quantum geometries of the normal state bands are the same when the superconductivity appears in the ferromagnetic and paramagnetic phases, the emerging superconducting phases have very different superfluid weights. In the case of superconductivity appearing on the spin-polarized bands the superfluid weight varies a lot as a function of model parameters, whereas in the absence of spin polarization the singlet-superconductivity has a robust and large superfluid weight originating from the quantum geometry. In order to better understand the effects caused by the coupling of the bands on the superfluid weight, we introduce a perturbation to our Hamiltonian which enables the realization of the  superconductivity on an isolated flat band, and study the behavior of the superfluid weight when the system is tuned from this ideal paradigmatic limit to the models proposed for copper-doped lead apatite.    

\section{Normal state Hamiltonian \label{normal-state}}

\begin{figure}
    \centering
    \includegraphics[width=0.95\linewidth]{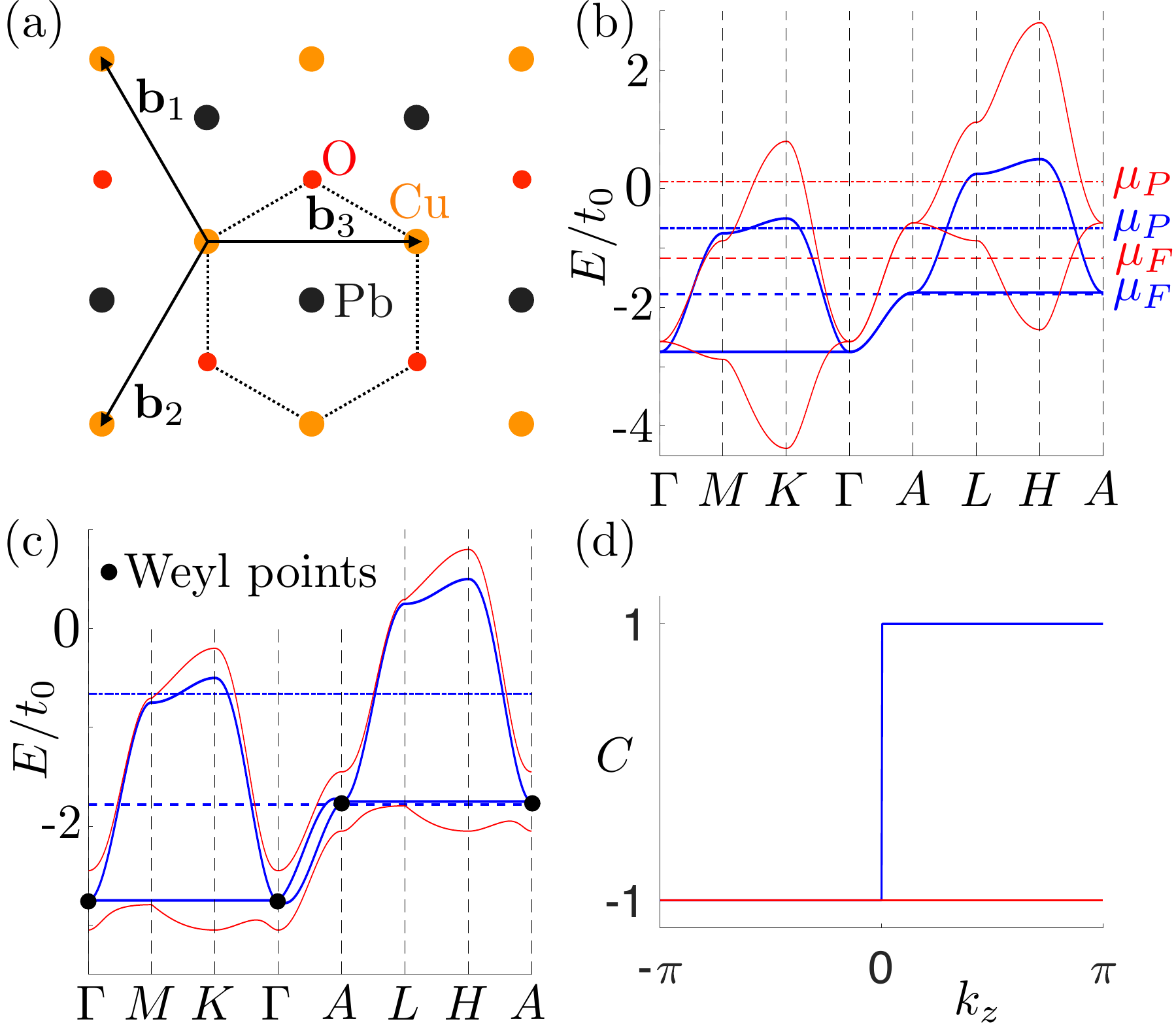}
    \caption{(a) Lattice structure showing the positions of the most relevant atoms for the low-energy theory. The flat bands originate mostly from the $d_{xz}$ and $d_{yz}$ orbitals of Cu atoms  forming a triangular lattice in $(x,y)$-planes. Together with the Pb atoms  the Cu atoms form honeycomb lattice, where all atoms are in the same layer. The O atoms, which are located in a different layer, form a buckled honeycomb lattice together with Cu atoms. The $p_x$ and $p_y$ orbitals of O atoms mediate hoppings between the Cu orbitals. The full 3D structure is obtained by stacking these layers.  (b) The band structure of the Hamiltonian (\ref{H_0}) with flat-band parameters (blue) and H-M parameters (red), given in Table \ref{table:tb_params}, along the high-symmetry path in the Brillouin zone. The Fermi levels in the ferromagnetic ($\mu_F$) and paramagnetic ($\mu_P$) phases are shown with blue and red dashed lines for the two sets of parameters. Here $\Gamma=(0,0,0)$, $M=(0, 2 \pi/3,0)$, $K=(2\pi/(3 \sqrt{3}),2\pi/3, 0)$, $A=(0,0,\pi)$, $L=(0, 2 \pi/3,\pi)$, $H=(2\pi/(3 \sqrt{3}),2\pi/3, \pi)$.  (c) Effects of mirror symmetry breaking hopping ($t_{z-}=t_0/12$, blue line) and spin-orbit coupling ($\lambda=0.3 t_0$, red line) on the band structure obtained with the flat-band parameters. The double Weyl points in the mirror symmetry broken case appear at $\Gamma$ and $A$. (d) Chern number of the lower band as a function of $k_z$ in the presence of mirror symmetry breaking term ($t_{z-}=t_0/12$, blue line) and spin-orbit coupling ($\lambda=0.3 t_0$, red line).}
    \label{fig:normal-state}
\end{figure}

\begin{table}[h!]
\centering
\begin{tabular}{|c | c |c |c|} 
 \hline
 \textbf{parameter} & \textbf{flat-band parameters} & \textbf{H-M parameters} \\ [0.5ex] 
 \hline
 $\bar{t}_0/t_0$ &  1  &0.7 \\ [0.5ex]
 \hline
  $\hat{t}_0/t_0$ & 1 &2.3 \\[0.5ex]
 \hline
 $t_z/t_0$     & 0.25 &0.5 \\[0.5ex]
 \hline
 $\mu_F/t_0$ & -1.78 & -1.17 \\[0.5ex]
 \hline
 \hline
 $\mu_P/t_0$ & -0.66 & 0.12 \\[0.5ex]
 \hline
 \end{tabular}
\caption{Table of parameter values used for the calculations in this paper. Here $\mu_F$ ($\mu_P$) is the value of the chemical potential in the ferromagnetic (paramagnetic) phase. The hopping parameter $t_0 \approx 30$ meV.}
\label{table:tb_params}
\end{table}

Our starting point is the low-energy Hamiltonian for $d_{xz}$ and $d_{yz}$ orbitals of Cu atoms forming a triangular lattice shown in Fig.~\ref{fig:normal-state}(a). It can be written as \cite{Hirchmann2023}
\begin{eqnarray}
H_0(\mathbf{k})&=&\sum_{\nu=0}^4 d_\nu(\mathbf{k}) \sigma_\nu, \nonumber \\
d_0(\mathbf{k})&=& \bar{t}_0 \bigg[-\frac{1}{4} \sum_{i=1}^3\cos (\mathbf{k} \cdot \mathbf{b}_i)  - 3/2\bigg]  - 2 t_z \cos k_z, \nonumber \\
d_1(\mathbf{k}) &=& \frac{\sqrt{3} t_0}{4} \bigg[\cos(\mathbf{k} \cdot \mathbf{b}_1) - \cos(\mathbf{k} \cdot \mathbf{b}_2) \bigg], \label{H_0}  \\
d_2(\mathbf{k}) &=& \frac{\sqrt{3} \hat{t}_0}{4} \sum_{i=1}^3 \sin (\mathbf{k} \cdot \mathbf{b}_i), \nonumber \\
d_3(\mathbf{k}) &=& \frac{t_0}{4} \bigg[  \cos(\mathbf{k} \cdot \mathbf{b}_1) + \cos(\mathbf{k} \cdot \mathbf{b}_2)  - 2\cos(\mathbf{k} \cdot \mathbf{b}_3) \bigg], \nonumber
\end{eqnarray}
where the hoppings between the Cu orbitals, $t_0$, $\hat{t}_0$, $\bar{t}_0$ and $t_z$, are mediated by the $p_x$ and $p_y$ orbitals of O atoms [see Fig.~\ref{fig:normal-state}(a)],  and  the lattice vectors $\mathbf{b}_i$ satisfy: $\mathbf{b}_1 = -\sqrt{3} \hat{e}_x/2+3 \hat{e}_y/2$, $\mathbf{b}_2 = -\sqrt{3} \hat{e}_x/2-3 \hat{e}_y/2$, $\mathbf{b}_3 = \sqrt{3} \hat{e}_x$. The chemical potential $\mu$ should be determined so that the filling of the bands is correct.
This nearest-neighbor hopping model can be obtained from symmetry considerations by requiring that the Hamiltonian satisfies spinless time-reversal symmetry (TRS)
\begin{equation}
H_0^T(-\mathbf{k})= H_0(\mathbf{k}), \label{spinless-TRS}
\end{equation}
a three-fold rotation symmetry 
\begin{equation}
U^\dag H_0(\mathbf{b}_1 \cdot \mathbf{k} \to \mathbf{b}_2 \cdot \mathbf{k} \to \mathbf{b}_3 \cdot \mathbf{k} \to \mathbf{b}_1 \cdot \mathbf{k} ) U=H_0(\mathbf{k}), \label{three-fold}
\end{equation}
where 
\begin{equation}
U=\begin{pmatrix} \cos (2 \pi/3) & \sin (2 \pi/3) \\
 -\sin (2 \pi/3) & \cos (2 \pi/3) \end{pmatrix},
\end{equation}
a spinless in-plane mirror symmetry
\begin{equation}
\sigma_z H_0(-k_x, k_y, k_z) \sigma_z = H_0(k_x, k_y, k_z), \label{in-plane}
\end{equation}
and an out-of-plane mirror symmetry 
\begin{equation}
H_0(k_x, k_y, - k_z) = H_0(k_x, k_y, k_z). \label{z-mirror}
\end{equation}
Two additional in-plane mirror symmetries can be obtained by combining (\ref{in-plane}) with the rotation symmetry.  If one considers only the dominant hopping paths along the bond directions the hoppings satisfy $t_0= \hat{t}_0=\bar{t}_0$. This was shown in Ref.~\cite{tavakol2023minimal} for hoppings mediated by the $d_{xz}$ and $d_{yz}$ orbitals of the Pb atoms, but it remains true also in a more realistic model where the hoppings are mediated by the $p_x$ and $p_y$ orbitals of O atoms. Interestingly, for this relation between the hopping amplitudes the energy of the lower band is given by \cite{tavakol2023minimal}
\begin{equation}
\varepsilon_1(\mathbf{k})=-\frac{9}{4}t_0 - 2t_z \cos k_z, 
\end{equation}
producing a perfectly flat band as a function of $k_x$ and $k_y$ \cite{tavakol2023minimal}. In Table \ref{table:tb_params}  we refer to these hopping amplitudes as the flat-band parameters, where we have additionally chosen $t_z=0.25 t_0$ in agreement with Ref.~\cite{tavakol2023minimal}. According to the DFT calculations \cite{griffin2023origin,si2023electronic, kurleto2023pbapatite, cabezasescares2023theoretical, jiang2023pb9cupo46oh2}, the hopping amplitudes of the copper-doped lead apatite do not satisfy this relationship, leading to weak dispersion as a function of $k_x$ and $k_y$, which is illustrated for the parameters proposed by  Hirschmann and Mitscherling \cite{Hirchmann2023} (see H-M parameters in Table \ref{table:tb_params}) in Fig.~\ref{fig:normal-state}(b). It should be emphasized that the energy scale is $t_0 \approx 30$ meV \cite{Hirchmann2023}, so that the bands are always relatively flat.   

There exist also two important corrections to the Hamiltonian given in Eq.~(\ref{H_0}). The O atoms break the mirror symmetries (\ref{in-plane}) and (\ref{z-mirror})  \cite{griffin2023origin,si2023electronic, kurleto2023pbapatite, cabezasescares2023theoretical, jiang2023pb9cupo46oh2, Hirchmann2023, lee2023effective}, resulting in a mirror-symmetry-breaking hopping term between the Cu orbitals \cite{Hirchmann2023} 
\begin{equation}
H_{mb}(\mathbf{k})=-2 t_{z-} \sin k_z \ \sigma_2. \label{mirror-breaking}
\end{equation}
Notice that this perturbation obeys the spinless TRS (\ref{spinless-TRS}) and three-fold rotational symmetry (\ref{three-fold}). Additionally, the spin-orbit coupling cannot be neglected because the unit cell contains large number of Pb atoms \cite{bai2023ferromagnetic} and also the Cu is known to have spin-orbit coupling comparable to tight-binding hopping parameters in this model \cite{Fabian17}. Similarly, as in the recent work \cite{zhou2023cusubstituted}, we introduce the spin-orbit coupling with Hamiltonian
\begin{equation}
H_{so}(\mathbf{k})= \lambda \sigma_2. \label{so}
\end{equation}
This Hamiltonian breaks the spinless TRS (\ref{spinless-TRS}) and in-plane mirror (\ref{in-plane}) symmetries. Although the spin-orbit coupling does not break the true TRS and in-plane mirror symmetries, the spinless versions of these symmetries are broken. From the physics viewpoint this means that in the spin-polarized phase there is no TRS or in-plane mirror symmetries because  the combination of spin-splitting field and spin-orbit coupling breaks these symmetries. In the paramagnetic phase this perturbation preserves spinful TRS  and spinful mirror symmetries (see below). 

The two types of symmetry breaking terms (\ref{mirror-breaking}) and (\ref{so}) lead to different kinds of momentum-space topologies of the normal state Hamiltonian. The Hamiltonian (\ref{mirror-breaking}) introduces double Weyl points at $\Gamma$ and $A$ \cite{Hirchmann2023} as shown in Fig.~\ref{fig:normal-state}(c), so that the Chern number, calculated for the lower band over the ($k_x, k_y$)-plane with fixed $k_z$, satisfies \cite{Hirchmann2023}
\begin{equation}
C=\begin{cases} 
-1, & -\pi < k_z < 0 \\
+1, & \hspace{0.29cm} 0 < k_z < \pi
\end{cases}
\label{spin-Chern-mb}
\end{equation}
as shown in Fig.~\ref{fig:normal-state}(d). Introducing the spin-orbit coupling term in the presence of mirror-breaking term moves the Weyl points away from $\Gamma$ and $A$ points so that they finally merge and annihilate each others at $\lambda = 2 t_{z-}$ \cite{zhou2023cusubstituted}. If $\lambda > 2 t_{z-}$ so that the Weyl points have been annihilated the Chern number satisfies 
\begin{equation}
C=-1, \textrm{ for all } k_z
\label{spin-Chern-s0}
\end{equation}
as shown in Fig.~\ref{fig:normal-state}(d).

\subsection{Ferromagnetic phase}

In the ferromagnetic phase, favoured by the DFT calculations \cite{griffin2023origin,si2023electronic, kurleto2023pbapatite, cabezasescares2023theoretical, jiang2023pb9cupo46oh2}, the two spin-$\uparrow$ flat bands are described by a Hamiltonian 
\begin{equation}
H_{\uparrow}(\mathbf{k})=H_{0}(\mathbf{k})+H_{mb}(\mathbf{k})+H_{so}(\mathbf{k})-\mu_F \, \sigma_0
\label{spin-polarized-H}
\end{equation}
and the spin-$\downarrow$ are far below the Fermi level and fully occupied, so that they do not influence the low-energy theory close to the Fermi level at all. Therefore, the spin-$\uparrow$ orbitals are on average half-filled, so that the total filling factor of the two bands of the Hamiltonian (\ref{spin-polarized-H}) is $1$.   This means that the Fermi level $\mu_F$ in the system is in-between the Weyl point energies as shown in Fig.~\ref{fig:normal-state}(c), so that the system is ideal for observing the Fermi arcs protected by the spin-Chern number $C_{\uparrow}=C$, where $C$ is given by Eq.~(\ref{spin-Chern-mb}). Importantly, the system remains topologically non-trivial if the spin-orbit coupling causes the annihilation of the Weyl points because the spin-Chern number is  $C_\uparrow=C=-1$ for all $k_z$, giving rise to surface states which resemble the Fermi arcs of the Weyl semimetal phase in the surface density of states but they have the same chirality for all values of $k_z$, so their effect on transport properties is expected to be different.  In principle, such kind of chiral currents may repel the magnetic field and cause diamagnetism even in the absence of superconductivity. This could happen because in the absence of magnetic field the sample will likely contain domains where the magnetization direction points randomly in both directions, and the external magnetic field can align the magnetization directions so that the chiral current expels the magnetic field.

\subsection{Paramagnetic phase}

Although the existence of the paramagnetic phase in this material is not support by the DFT calculations, we consider it in this work because the ferromagnetism and superconductivity could be competing orders, and the transition to the superconducting state could be accompanied by the disappearance of spin polarization similarly as in other systems \cite{Annica2017, Ojajarvi2018, Lau21, Hu2020}. In the paramagnetic phase 
\begin{equation}
H(\mathbf{k})=\begin{pmatrix} 
H_\uparrow (\mathbf{k}) & 0 \\
0 & H_{\downarrow} (\mathbf{k})
\end{pmatrix},
\label{paramagnetic-H}
\end{equation}
where
\begin{equation}
H_{\uparrow}(\mathbf{k})=H_{0}(\mathbf{k})+H_{mb}(\mathbf{k})+H_{so}(\mathbf{k})-\mu_P \, \sigma_0 \\
\label{paramagnetic-Hup}
\end{equation}
and
\begin{equation}
H_{\downarrow}(\mathbf{k})=H_{\uparrow}^T(-\mathbf{k}). \\
\label{paramagnetic-Hdown}
\end{equation}
The topological properties and the energy-momentum dispersions of the bands are essentially the same as in ferromagnetic phase because the low-energy theory consist of $d_{xz}$ and $d_{yz}$ orbitals so that the atomic spin-orbit coupling does not couple the two spin blocks. Therefore we obtain $4$ similar flat bands as discussed above. The spin-Chern number $C_\uparrow(k_z)$ of the lower flat band of $H_{\uparrow}(\mathbf{k})$ as a function of $k_z$ behaves similarly as in the ferromagnetic phase and the spin-Chern number $C_\downarrow(k_z)$ of the lower flat band of $H_{\downarrow}(\mathbf{k})$ satisfies 
\begin{equation}
C_\downarrow(k_z)=-C_\uparrow(-k_z).
\end{equation}
However, the important difference to ferromagnetic case is that the spin-splitting field is now absent, so that the two flat bands in each spin sector must be $3/4$ filled on average in order that the total filling of the four bands is $3$. This means that the Fermi level is significantly above the Weyl points [see Fig.~\ref{fig:normal-state}(c)] and the surface states are significantly below the Fermi level.

In addition to the symmetries discussed above, in the paramagnetic phase the Hamiltonian (\ref{paramagnetic-H}) satisfies spinful TRS 
\begin{equation}
s_y \sigma_0 H^T(-\mathbf{k}) s_y \sigma_0 = H(\mathbf{k})
\end{equation}
and spin-rotation symmetry around $z$-axis
\begin{equation}
s_z \sigma_0 H(\mathbf{k}) s_z \sigma_0 = H(\mathbf{k}),
\end{equation}
where we have denoted the Pauli matrices in the spin space with $s_i$ and in the orbital space with $\sigma_i$. If $t_{z-}=0$, the Hamiltonian satisfies also spinful mirror symmetry
\begin{equation}
s_x \sigma_z H(-k_x, k_y, k_z) s_x \sigma_z =  H(k_x, k_y, k_z).
\end{equation}

\section{Minimal models for the superconducting phase}

\subsection{Superconductivity in the spin-polarized phase}

Since we are looking for the possibility of high-temperature superconductivity we assume momentum-independent on-site pairing order parameter. 
In the spin-polarized phase the only possible order parameter, satisfying these requirements, is the interorbital order parameter which pairs the electrons in the $d_{xz}$ and $d_{yz}$ orbitals \cite{tavakol2023minimal}. Therefore, we assume that the Bogoliubov-de Gennes (BdG) Hamiltonian in the spin polarized superconducting phase is 
\begin{equation}
H_{BdG}(\mathbf{k})=\begin{pmatrix}
H_{\uparrow}(\mathbf{k}) & \Delta \\
\Delta^\dag & -H^T_{\uparrow}(-\mathbf{k})
\end{pmatrix},
\label{BdG-polarized}
\end{equation}
where 
\begin{equation}
\Delta=\begin{pmatrix} 0 & \Delta_0 \\
       -\Delta_0 & 0  \end{pmatrix}.
\end{equation}
Because the scale $t_0$ of the hopping parameters in these models is small, and we are looking of the possibility of high-temperature superconductivity, the parameter $\Delta_0$ can be comparable to $t_0$.

\begin{figure}
    \centering
    \includegraphics[width=0.92\linewidth]{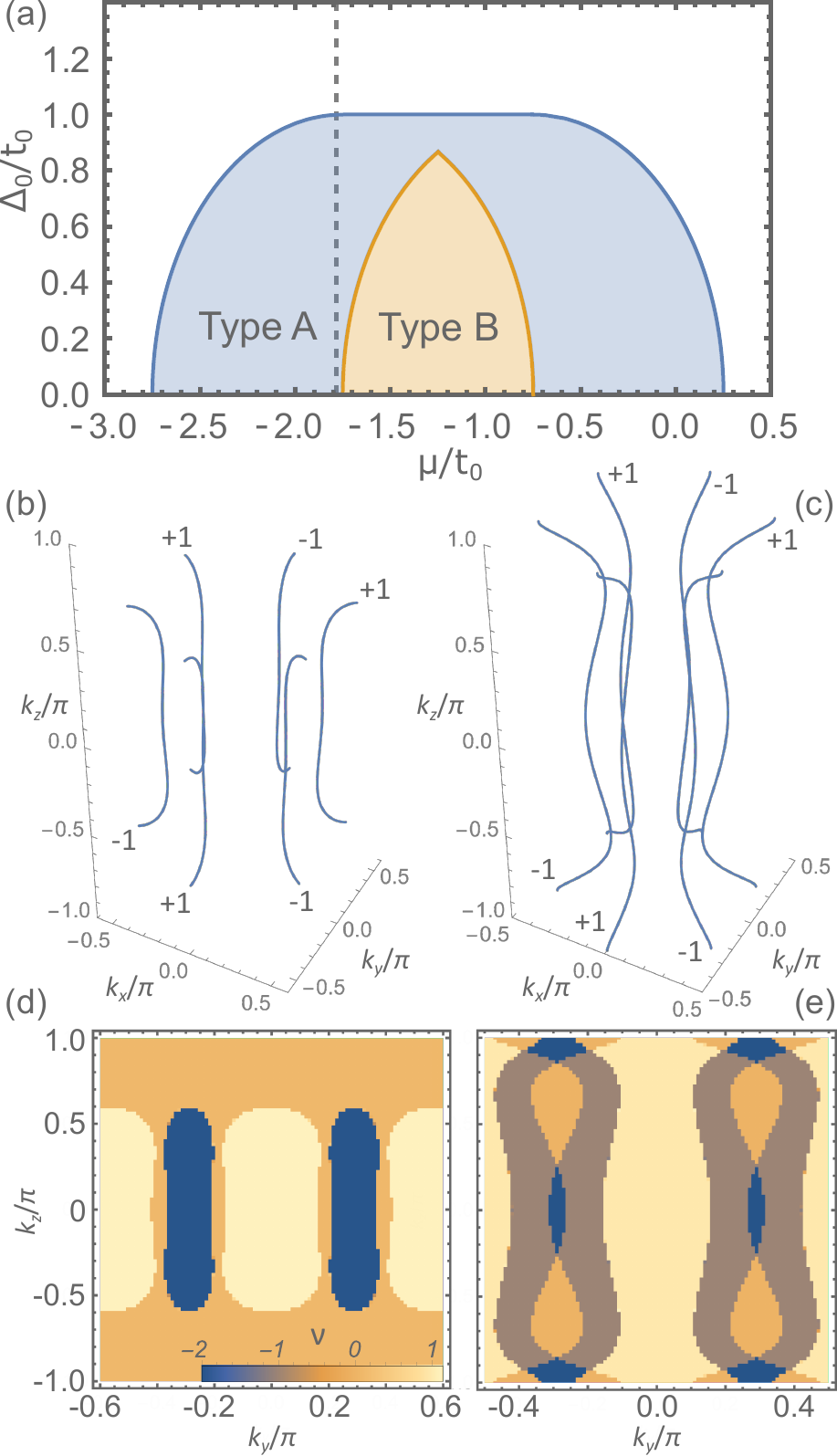}
    \caption{(a) Phase diagram of the spin-polarized  superconducting model (\ref{BdG-polarized})  as a function of $\mu$ and $\Delta$ for $t_{z-}=\lambda=0$ and other parameters given by the flat-band parameters in Table \ref{table:tb_params}. The expected chemical potential in the ferromagnetic phase $\mu=\mu_F$ is shown with a dashed line. The phase diagram contains a topologically trivial fully gapped phase, and Type A (nodal loops) and Type B (nodal lines going through the Brillouin zone in $k_z$ direction) gapless topological phases. (b) Illustration of the nodal lines in the momentum space for the Type A gapless phase. (c) Same for the Type B gapless phase. (d) Winding number $\nu(k_y, k_z)$ determining the number of surface flat bands at the side surfaces as a function surface momentum $(k_y, k_z)$ in Type A gapless phase. (e) Same for Type B gapless phase. In (b) and (d) the parameters are $\mu= -1.7$ and  $\Delta=0.8$.  In (c) and (e) the parameters are $\mu= -1.7$ and  $\Delta=0.3$.  
    \label{fig:spin-polarized-SC}
    }
\end{figure}

\begin{figure}
    \centering
    \includegraphics[width=0.92\linewidth]{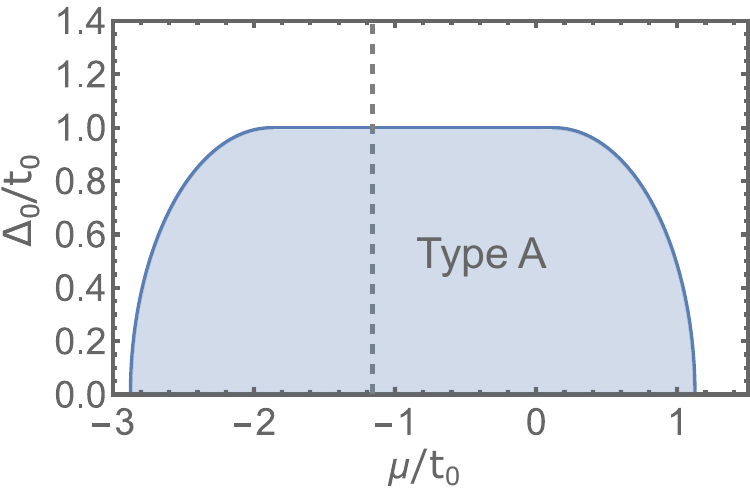}
    \caption{Phase diagram of the spin-polarized  superconducting model (\ref{BdG-polarized})  as a function of $\mu$ and $\Delta$ for $t_{z-}=\lambda=0$ and other parameters given by the H-M parameters in Table \ref{table:tb_params}. The expected chemical potential in the ferromagnetic phase $\mu=\mu_F$ is shown with a dashed line. In this case the phase diagram contains only the trivial phase and the Type A gapless phase. 
    \label{fig:spin-polarized-SC-H-M}
    }
\end{figure}

\begin{figure}
    \centering
    \includegraphics[width=0.95\linewidth]{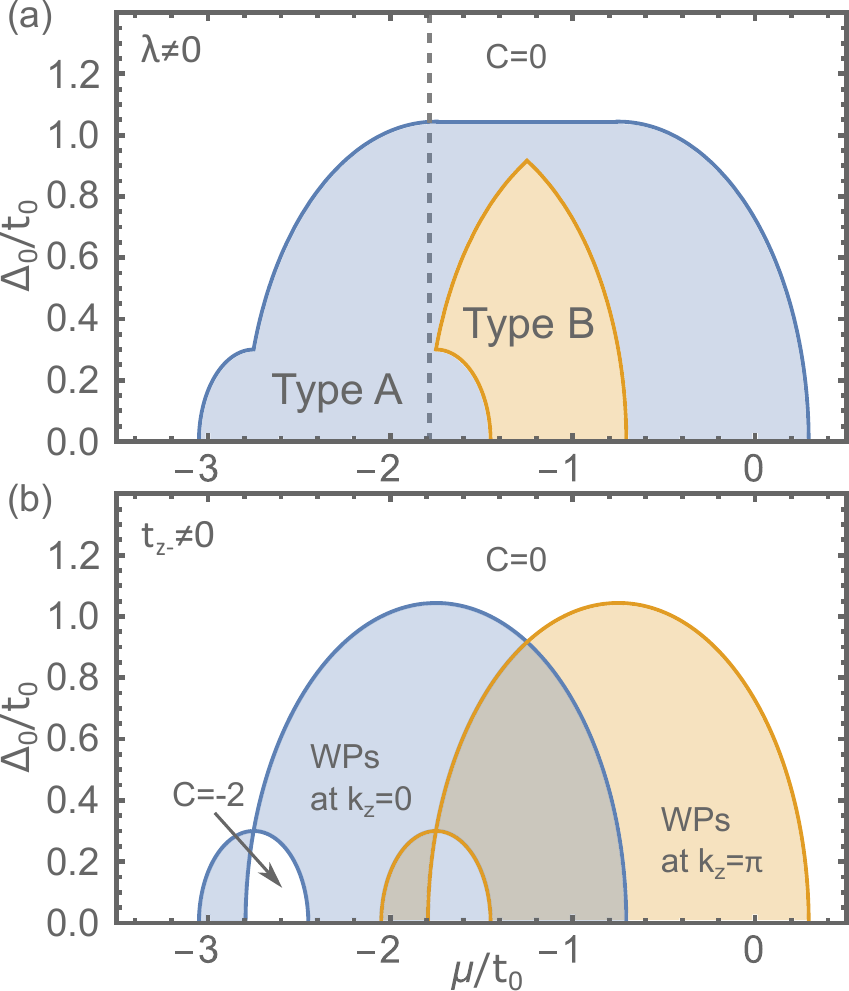}
    \caption{Phase diagrams of the spin-polarized  superconducting model (\ref{BdG-polarized})  as a function of $\mu$ and $\Delta$ for the flat-band parameters given in Table \ref{table:tb_params} in presence of symmetry-breaking terms. (a) In the presence of spin-orbit coupling $\lambda=0.3 t_0$, the phase diagrams contains Type A and Type B gapless phases protected by the combination of the symmetries (\ref{PHS}) and (\ref{1D-inversion}). (b) In the presence of spin-orbit coupling $\lambda=0.3 t_0$ and  mirror-symmetry breaking $t_{z-}=t_0/12$ the symmetry (\ref{1D-inversion}) is valid only at $k_z=0, \ \pi$. The protected Weyl points (WPs) at these planes are shown with blue and orange colors. The $C=-2$ indicates a region, where the two lowest bands of the BdG Hamiltonian (\ref{BdG-polarized}) carry a Chern number $C=-2$ at $k_z=0$ plane. Therefore, it is topologically distinct from the trivial $C=0$ phase.
    \label{fig:spin-polarized-SC2}
    }
\end{figure}

In the absence of spin-orbit coupling ($\lambda=0$) the Hamiltonian (\ref{BdG-polarized}) can be written as
\begin{equation}
H_{BdG}(\mathbf{k}) = H_{\uparrow}(\mathbf{k}) \tau_z-\Delta_0 \tau_y \sigma_y, 
\end{equation}
where we have denoted the Pauli matrices in the Nambu space with $\tau_i$. This Hamiltonian 
satisfies a particle-hole symmetry (PHS)
\begin{equation}
\tau_x H^T_{BdG}(-\mathbf{k}) \tau_x= -H_{BdG}(\mathbf{k}) \label{PHS}
\end{equation}
and time-reversal symmetry 
\begin{equation}
H^T_{BdG}(-\mathbf{k}) = H_{BdG}(\mathbf{k})
\end{equation}
so that there exists a chiral symmetry
\begin{equation}
\tau_x H_{BdG}(\mathbf{k}) \tau_x = - H_{BdG}(\mathbf{k}).
\end{equation}
This means that the Hamiltonian can be transformed into a  block-off-diagonal form
\begin{equation}
U^\dag H_{BdG}(\mathbf{k}) U =\begin{pmatrix}
0 & D(\mathbf{k}) \\ 
D^\dag(\mathbf{k}) &0
\end{pmatrix},
\end{equation}
where $U=(\tau_z+\tau_x)/\sqrt{2}$ and $D(\mathbf{k})=H_{\uparrow}(\mathbf{k}) - \Delta$. In the presence of chiral symmetry 3D Hamiltonians can support nodal lines, which are protected by the winding number of $z(\mathbf{k})={\rm Det} D(\mathbf{k})/|{\rm Det} D(\mathbf{k})|$ when one goes around the nodal line in the momentum space \cite{RevModPhys.88.035005}. Figure \ref{fig:spin-polarized-SC}(a) shows that the phase diagram of the model (\ref{BdG-polarized}) with the flat-band parameters, given in Table \ref{table:tb_params}, and $t_{z-}=0$ contains a fully gapped topologically trivial superconducting phase and two different kinds of gapless topological phases. In Type A gapless phase the nodal lines form loops inside the Brillouin zone and in Type B gapless phase the nodal lines go through the Brillouin zone in $k_z$ direction \cite{Hyart2018, PhysRevB.96.155105, bouhon2017bulk}, as illustrated in Fig.~\ref{fig:spin-polarized-SC}(b),(c). In both cases the nodal lines give rise to surface flat bands. The number of zero-energy surface states at the surface momentum $(k_l, k_m)$ is determined by winding number of $z(\mathbf{k})$ across the Brillouin zone in the third momentum direction $k_n$, as illustrated for surfaces parallel to $(y,z)$-plane in Fig.~\ref{fig:spin-polarized-SC}(d),(e). The mirror symmetry breaking term $H_{mb}$ preserves the chiral symmetry and therefore the phase diagram is qualitatively similar also when $t_{z-} \ne 0$. However, we point out that depending on the values of the model parameters the regions of the different phases in the $(\mu, \Delta)$-plane can look very different. For example for the H-M parameters only the trivial phase and Type A gapless phase are present as shown in Fig.~\ref{fig:spin-polarized-SC-H-M}.

The presence of spin-orbit coupling ($\lambda \ne 0$) breaks the chiral symmetry. Nevertheless, it follows from the symmetries discussed in Section \ref{normal-state} and from the fact that spin-orbit coupling (\ref{so}) is momentum-independent that 
\begin{equation}
H_{BdG}(0, -k_y, k_z)=H_{BdG}(0, k_y, k_z). \label{1D-inversion}
\end{equation}
Together with the PHS (\ref{PHS}) this guarantees that there exists a $\mathbb{Z}_2$-Pfaffian invariant \cite{Z21, Z22}, which can give rise to topologically protected band crossings in the plane $k_x=0$ (and in the other planes related to this by the three-fold rotational symmetry). We indeed find that this kind of topological nodal lines exist in a large part of the parameter space as shown in Fig.~\ref{fig:spin-polarized-SC2}(a). When both spin-orbit coupling ($\lambda \ne 0$) and mirror symmetry breaking ($t_{z-} \ne 0$) are present the symmetry (\ref{1D-inversion}) exist only in the planes $k_z=0, \ \pi$. Therefore, it no longer protects nodal lines but instead we obtain topologically protected Majorana-Weyl points at the high-symmetry lines as shown in Fig.~\ref{fig:spin-polarized-SC2}(b). Even in the absence of the Weyl points the model can support topologically distinct phases because the Chern number calculated for the two lowest bands of the BdG Hamiltonian (\ref{BdG-polarized}) in the $k_z=0$ plane can have values $-2$ or $0$ as shown in Fig.~\ref{fig:spin-polarized-SC2}.

We point out that in Fig.~\ref{fig:spin-polarized-SC2} we have only characterized the phases in terms of the topologically protected band crossings at the high-symmetry planes and lines discussed above, but there can be gap closings also in other parts of the Brillouin zone. In particular,  the phases are expected to be gapless in some regions of the parameter space because the breaking of the chiral symmetry enables the possibility of indirect gap closings. 

\subsection{Superconductivity in the paramagnetic phase}

In the case of paramagnetic phase the most likely candidate for momentum-independent order parameter is the $s$-wave singlet pairing. Therefore, the BdG Hamiltonian in this case is
\begin{equation}
H_{BdG}(\mathbf{k})=\begin{pmatrix}
H_{\uparrow}(\mathbf{k}) & \Delta_0 \sigma_0 \\
\Delta_0 \sigma_0 & -H^T_{\downarrow}(-\mathbf{k})
\end{pmatrix}.
\label{BdG-s-wave}
\end{equation}
Because of the spinful TRS $H^T_{\downarrow}(-\mathbf{k})=H_{\uparrow}(\mathbf{k})$
this Hamiltonian satisfies a chiral symmetry
\begin{equation}
\tau_y H_{BdG}(\mathbf{k}) \tau_y = - H_{BdG}(\mathbf{k}).
\end{equation}
For $\Delta_0 \ne 0$ this Hamiltonian is always fully gapped and topologically trivial.

\section{Superfluid weight}

Using linear response theory the in-plane supercurrent response to the gradient of the phase can be expressed as
\begin{equation}
j_i=\frac{g_s}{2} \frac{\hbar}{2e} [D_s]_{ij} (\partial_j \phi-\frac{2e}{\hbar} A_j),
\end{equation}
where due to the three-fold rotational symmetry $[D_s]_{xy}=[D_s]_{yx}=0$, $[D_s]_{xx}=[D_s]_{yy}=D_s$, and
\begin{widetext}
\begin{eqnarray}
D_s&=&\frac{e^2}{\hbar^2}\frac{1}{V} \sum_{\mathbf{k}, i, j} \frac{n_F(E_j(\mathbf{k}))-n_F(E_i(\mathbf{k}))}{E_i(\mathbf{k})-E_j(\mathbf{k})} \bigg\{ \bigg| \langle \psi_i(\mathbf{k}) | \frac{\partial H_{BdG}(\mathbf{k})}{\partial k_x} |\psi_j(\mathbf{k})\rangle \bigg|^2  -\bigg|\langle \psi_i(\mathbf{k}) | \frac{\partial H_{BdG}(\mathbf{k})}{\partial k_x} \tau_z | \psi_j(\mathbf{k}) \rangle \bigg|^2 \bigg\}.
\label{Ds}
\end{eqnarray}
\end{widetext}
Here the degeneracy factor $g_s=1$ in the spin-polarized model (\ref{BdG-polarized}) and $g_s=2$ in the case of time-reversal invariant singlet superconducting model (\ref{BdG-s-wave}), $V$ is the volume of the sample, and $|\psi_i(\mathbf{k})\rangle$ and $E_i(\mathbf{k})$ are the eigenstates and eigenenergies of $H_{BdG}(\mathbf{k})$, respectively. In the derivation of Eq.~(\ref{Ds}) we utilized also the momentum-independence of the superconducting order parameter. In our numerical results we express the superfluid weight in units $D_0=e^2 t_0/\hbar^2$.
This is the natural choice of unit, because in these units the conventional contribution to the superfluid weight, $D_{s, conv} \approx e^2 n/m^*$, in the middle of the upper weakly dispersive band in the singlet superconducting phase is expected to be on the order $D_{s, conv}/D_0 \sim 1$. All the shown results are calculated at zero temperature.

It is useful to decompose $D_s$ to the conventional contribution determined by the single-particle spectrum (analogous to the conventional single-band case discussed above) and the contribution determined by the quantum geometry of the bands (interband contributions) \cite{Peotta2015, Liang2017, Hu2019,Fang2020,Julku2020,Hu2020, Rossi2021,Torma2021}. For this purpose we express the superfluid weight in terms of the eigenstates of the normal-state Bloch Hamiltonian $H_{BdG}(\mathbf{k}, \Delta=0)$. In the spin-polarized case (\ref{BdG-polarized}), we decompose the BdG eigenstates as
\begin{equation}
    |\psi_i\rangle = \sum^N_{m=1}\bigg[w_{+,im} \begin{pmatrix} |m(\mathbf{k})\rangle \\ 0 \end{pmatrix}+w_{-,im}\begin{pmatrix} 0 \\ |m(-\mathbf{k})\rangle^* \end{pmatrix} \bigg], \label{band-ansatz}
\end{equation}
where $|m(\mathbf{k})\rangle$ is the eigenvector of $H_\uparrow(\mathbf{k})$ with eigenvalue $\varepsilon_{m}(\mathbf{k})$ and $|m(-\mathbf{k})\rangle^*$ is the eigenvector of $H_\uparrow^T(-\mathbf{k})$ with eigenvalue $\varepsilon_{m}(-\mathbf{k})$. In the case of time-reversal invariant singlet superconductor (\ref{BdG-s-wave}), we perform the same expansion except that we replace $|m(-\mathbf{k})\rangle^*$ with the eigenvectors $|m(\mathbf{k})\rangle$ of $H_\downarrow^T(-\mathbf{k})=H_\uparrow(\mathbf{k})$ with eigenvalue $\varepsilon_{m}(\mathbf{k})$. By inserting this decomposition into Eq.~(\ref{Ds}) and keeping only the intraband contributions, we obtain the following expression for the conventional part of the superfluid weight  
\begin{widetext}
\begin{eqnarray}
    D^s_{conv} &=& -\frac{e^2}{\hbar^2}\frac{4}{V}\sum_{\mathbf{k},ijmn} \frac{n(E_j(\mathbf{k}))-n(E_i(\mathbf{k}))}{E_i(\mathbf{k})-E_j(\mathbf{k})}  
    w_{+,im}^* w_{+,jm} w_{-,jn}^* w_{-,in} M_{mn}(\mathbf{k}),
\end{eqnarray}
\end{widetext}
where in the spin-polarized case (\ref{BdG-polarized})
\begin{equation}
M_{mn}(\mathbf{k})= \partial_{k_x} \varepsilon_{m}(\mathbf{k}) \partial_{k_x} \varepsilon_{n}(-\mathbf{k}) 
\end{equation}
and in the singlet superconducting case (\ref{BdG-s-wave})
\begin{equation}
M_{mn}(\mathbf{k})= \partial_{k_x} \varepsilon_{m}(\mathbf{k}) \partial_{k_x} \varepsilon_{n}(\mathbf{k}). 
\end{equation}
The geometric contribution is obtained as 
\begin{equation}
 D^s_{geom}= D^s- D^s_{conv}.
\end{equation}

The standard paradigm of flat-band superconductivity is the time-reversal invariant $s$-wave singlet superconductivity in a system with spin-rotation symmetry around $z$-axis supporting a well-isolated flat band \cite{Peotta2015, Liang2017}. In this case $D_{s, conv}=0$ and 
\begin{equation}
D_{s,geom} = \frac{8e^2}{\hbar^2}  \Delta_0\,\sqrt{\nu(1-\nu)} \int\frac{d^d k}{(2\pi)^d}\: g_{xx}(\mathbf{k}),
\label{flat-band-geom-ideal}
\end{equation}
where $\nu$ the band filling and $d$ the dimension of the system.
The quantum metric $g_{\mu\nu}(\mathbf{k})$ is given by the real part of the quantum geometric tensor 
\begin{equation}
\mathcal{B}_{\mu\nu}(\mathbf{k}) = \langle \partial_{k_\mu} n(\mathbf{k}) | \big(1-|n(\mathbf{k})\rangle \langle n(\mathbf{k})| \big)\partial_{k_\nu} n(\mathbf{k})\rangle,
\end{equation}
where $|n(\mathbf{k}) \rangle$ is the normal state Bloch wave function of the flat band. 
The imaginary part of  $\mathcal{B}_{xy}(\mathbf{k})$ is $\Omega(\mathbf{k})/2$, where $\Omega(\mathbf{k})$ is the  Berry curvature.
Because $\mathcal{B}_{\mu\nu}(\mathbf{k})$ is positive semidefinite, in a two-dimensional system (or in a plane with fixed $k_z$ in a three-dimensional system) 
\begin{equation}
\int d^2k\: g_{xx}(\mathbf{k}) = \int d^2k\: g_{yy}(\mathbf{k}) \geq \frac{1}{2}\int  d^2k\: |\Omega(\mathbf{k})| \geq \pi|C|,
\end{equation}
where  $C = \frac{1}{2\pi}\int  d^2k\: \Omega(\mathbf{k})$ is the Chern number.
This means that if this type of system supports spin-up and spin-down flat bands carrying opposite spin-Chern numbers there exists 
a lower bound for $D_s$ determined by the absolute  value of the Chern number \cite{Peotta2015}.

As discussed in Section \ref{normal-state} the normal state Hamiltonians considered in this paper support nontrivial Chern numbers for large ranges of $k_z$ values. Therefore, one might be tempted to conclude that the superfluid weight has a lower bound in these systems. This conclusion is, however, wrong, because the models do not satisfy the   
assumptions required for the existence of the universal lower bounds. In the case of spin-polarized superconductivity (\ref{BdG-polarized}) the assumptions about singlet superconductivity and time-reversal symmetry are not satisfied. Additionally, in both models (\ref{BdG-polarized}) and (\ref{BdG-s-wave}) the bands are not well-isolated and they are not perfectly flat. In the following we discuss how these deviations from the idealized model influence the superfluid weight. In order to better understand our results, we write the Hamiltonian $H_{\uparrow}(\mathbf{k})$ as 
\begin{equation}
H_\uparrow(\mathbf{k})=\sum_{\mu=0}^4 d^\uparrow_\mu(\mathbf{k}) \sigma_\mu,
\end{equation}
where $d^\uparrow_0(\mathbf{k})=d_0(\mathbf{k})-\mu$, $d^\uparrow_1(\mathbf{k})=d_1(\mathbf{k})$, $d^\uparrow_2(\mathbf{k})=d_2(\mathbf{k})-2 t_{z-} \sin k_z + \lambda$, $d^\uparrow_3(\mathbf{k})=d_3(\mathbf{k})$, and $d_\mu(\mathbf{k})$ are given in Eq.~(\ref{H_0}).   
Then we introduce a perturbation to our Hamiltonian
\begin{equation}
\tilde{H}_{\uparrow}(\mathbf{k})= \xi
\begin{pmatrix} 
 \frac{|d^\uparrow(\mathbf{k})|- d_z^\uparrow(\mathbf{k})}{2 |d^\uparrow(\mathbf{k})| } & -\frac{d_x^\uparrow(\mathbf{k})-i d_y^\uparrow(\mathbf{k})}{2 |d^\uparrow(\mathbf{k})|} \\
 -\frac{d_x^\uparrow(\mathbf{k})+i d_y^\uparrow(\mathbf{k})}{2 |d^\uparrow(\mathbf{k})|} & \frac{|d^\uparrow(\mathbf{k})|+ d_z^\uparrow(\mathbf{k})}{2 |d^\uparrow(\mathbf{k})| }
\end{pmatrix},\label{perturbation}
\end{equation}
where $|d^\uparrow(\mathbf{k})|^2=d_x^\uparrow(\mathbf{k})^2+d_y^\uparrow(\mathbf{k})^2+d_z^\uparrow(\mathbf{k})^2$. This perturbation does not modify the eigenstates $|n(\mathbf{k})\rangle$ of the normal state Hamiltonian but shifts the lower band in energy by a constant $\xi$. Therefore, by controlling the magnitude of $\xi$ in this perturbation we can tune our model from the ideal case of an isolated flat band to the actual models proposed for copper-doped lead apatite.

\subsection{Superfluid weight in the superconducting spin-polarized phase}

\begin{figure}
    \centering
    \includegraphics[width=0.95\linewidth]{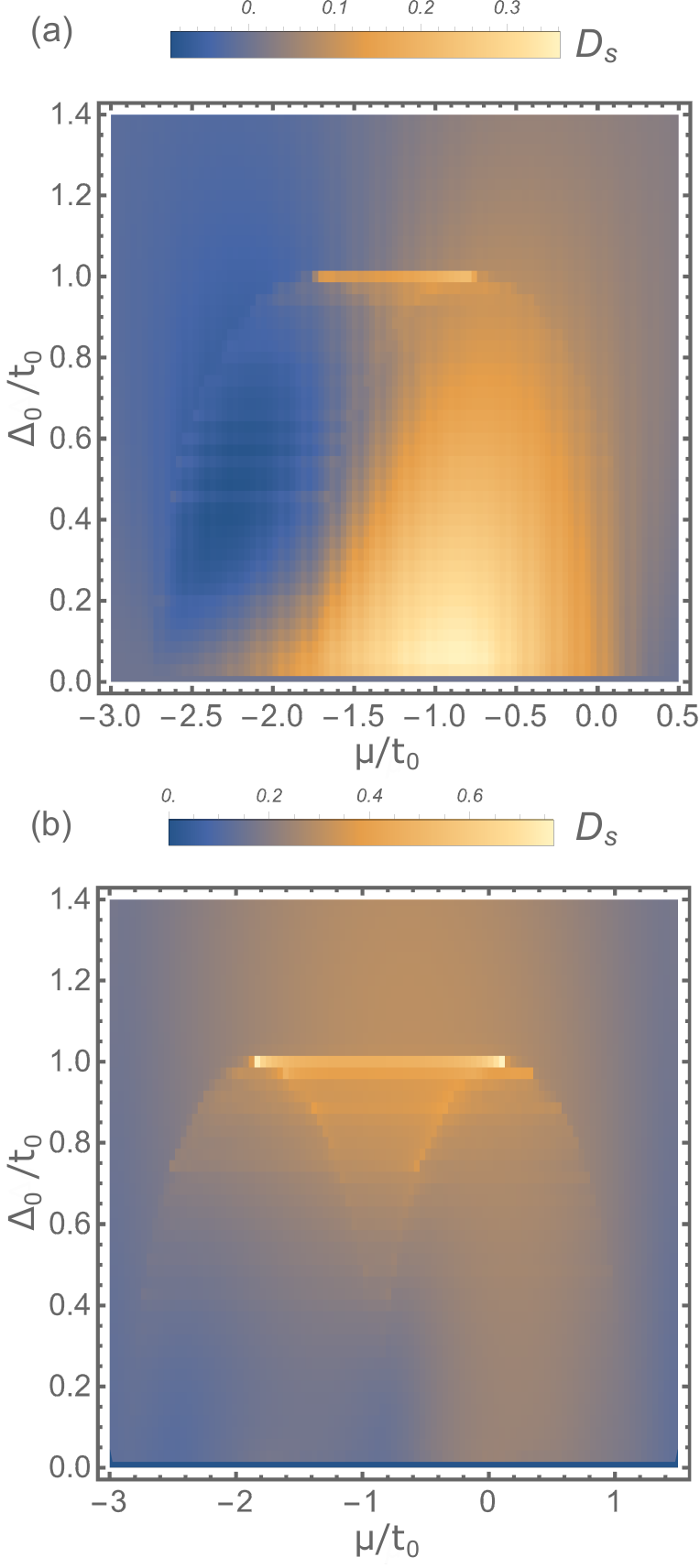}
    \caption{(a) Superfluid weight as a function of $\mu$ and $\Delta_0$ for the superconducting model (\ref{BdG-polarized}) with  $t_{z-}=\lambda=0$ and other parameters given by the flat-band parameters in Table \ref{table:tb_params}. (b) Same for the H-M parameters in Table \ref{table:tb_params}. 
}
    \label{fig:polarized-superfluidity}
\end{figure}

\begin{figure}
    \centering
    \includegraphics[width=0.9\linewidth]{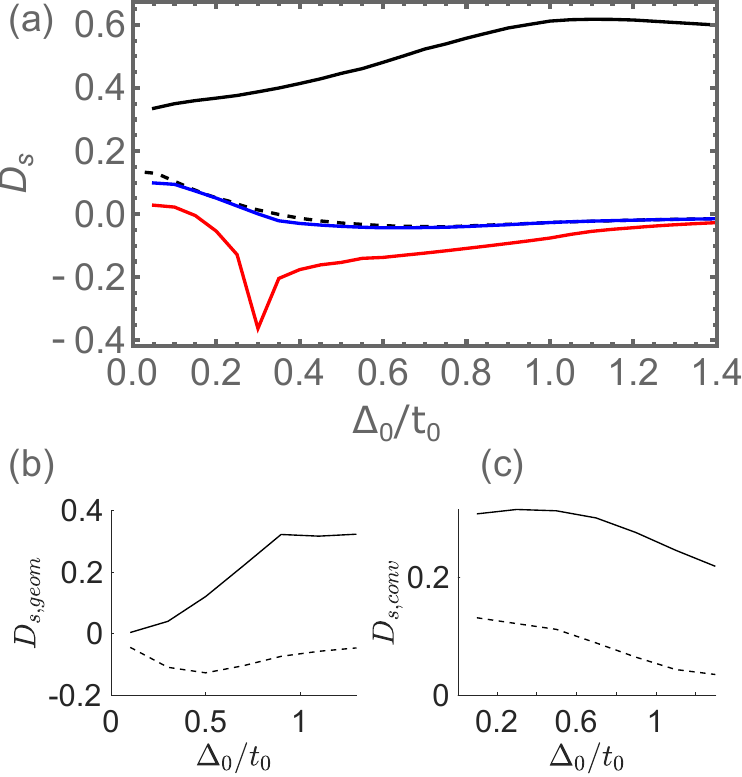}
    \caption{(a) Superfluid weight $D_s$ as a function of $\Delta_0$ for $\mu=\mu_F$. The black (dashed) line shows $D_s$ for the H-M (flat-band parameters) in   Table \ref{table:tb_params}. The blue (red) line shows the effect of mirror symmetry breaking hopping $t_{z-}=t_0/12$ (spin-orbit coupling $\lambda=0.3 t_0$) when the other parameters are given by the flat-band parameters. (b),(c) $D_{s, geom}$ and $D_{s, conv}$ as a function of $\Delta_0$ for the H-M (solid line) and flat-band (dashed line) parameters in Table \ref{table:tb_params}.}
    \label{fig:polarized-superfluidity-linecuts}
\end{figure}

Although the normal state bands support nontrivial topology and quantum geometry, we find that in the spin-polarized case there is no lower bound for $D_s$. As shown in Fig.~\ref{fig:polarized-superfluidity}(a) the superfluid weight can be even negative for a wide range of $\mu$ and $\Delta_0$ values, indicating instability of the spatially homogeneous superconducting state. Generically, we find that $D_s$ varies a lot depending on the values of the model parameters. For H-M parameters we obtain much larger values of $D_s$, $D_{s, geom}$ and $D_{s, conv}$ than for the flat-band parameters as shown Figs.~\ref{fig:polarized-superfluidity}(a)(b) and Fig.~\ref{fig:polarized-superfluidity-linecuts}. On the other hand, realistic values of mirror symmetry breaking hopping ($t_z=t_0/12$) and spin-orbit coupling ($\lambda=0.3 t_0$) do not influence $D_s$ significantly [see Fig.~\ref{fig:polarized-superfluidity-linecuts}(a)] even though they have significant effects on the topologies of the normal and superconducting states.

\subsection{Superfluid weight in the time-reversal invariant singlet $s$-wave superconducting phase}

\begin{figure}
    \centering
    \includegraphics[width=0.95\linewidth]{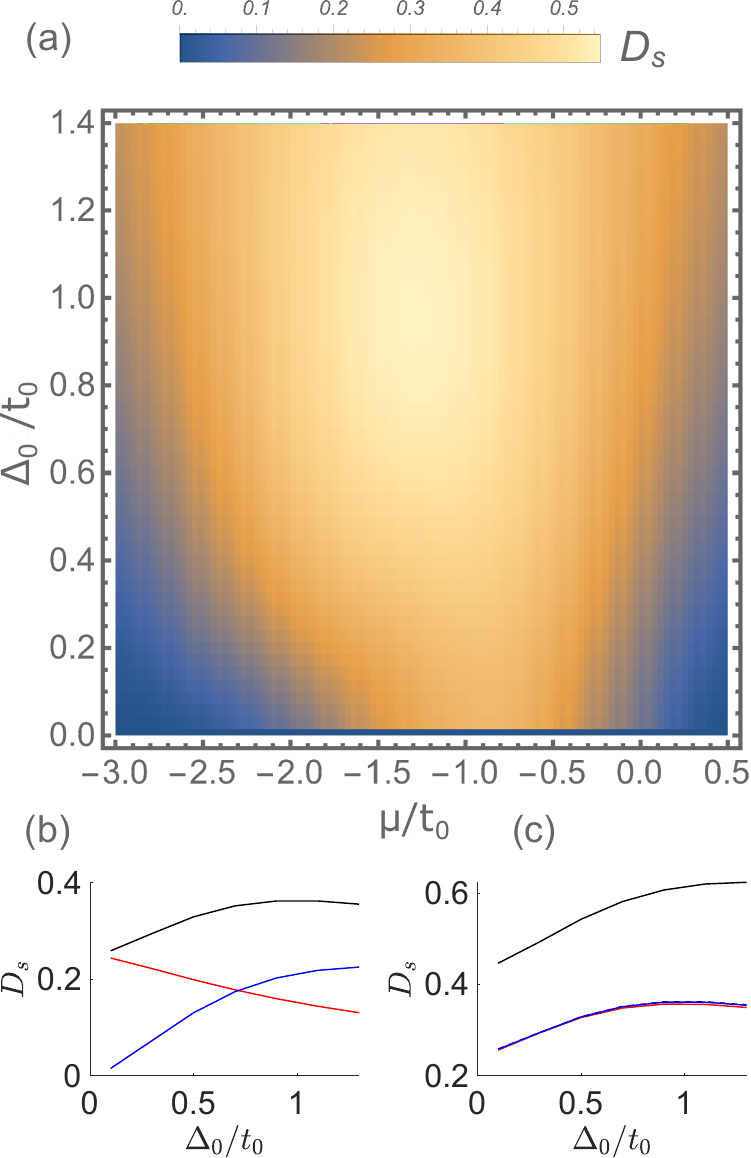}
\caption{(a) Superfluid weight as a function of $\mu$ and $\Delta_0$ for the superconducting model (\ref{BdG-s-wave}) with  $t_{z-}=\lambda=0$ and other parameters given by the flat-band parameters in Table \ref{table:tb_params}. (b) $D_{s}$ (black), $D_{s, conv}$ (red) and $D_{s, geom}$ (blue) as a function of $\Delta_0$ for $\mu=\mu_P$ and otherwise the same parameters. (c) The spin-orbit coupling $\lambda=0.3 t_0$ (red line) and the mirror-symmetry breaking hopping $t_{z-}=t_0/12$, (blue line) do not influence the results significantly. The superfluid weight with flat-band parameters and $\lambda=t_{z-}=0$ is shown with dashed line as a reference. On the other hand, $D_s$ is larger for H-M parameters in Table \ref{table:tb_params} (solid black line).}
    \label{fig:singlet-superfluidity}
\end{figure}

\begin{figure}
    \centering
    \includegraphics[width=0.95\linewidth]{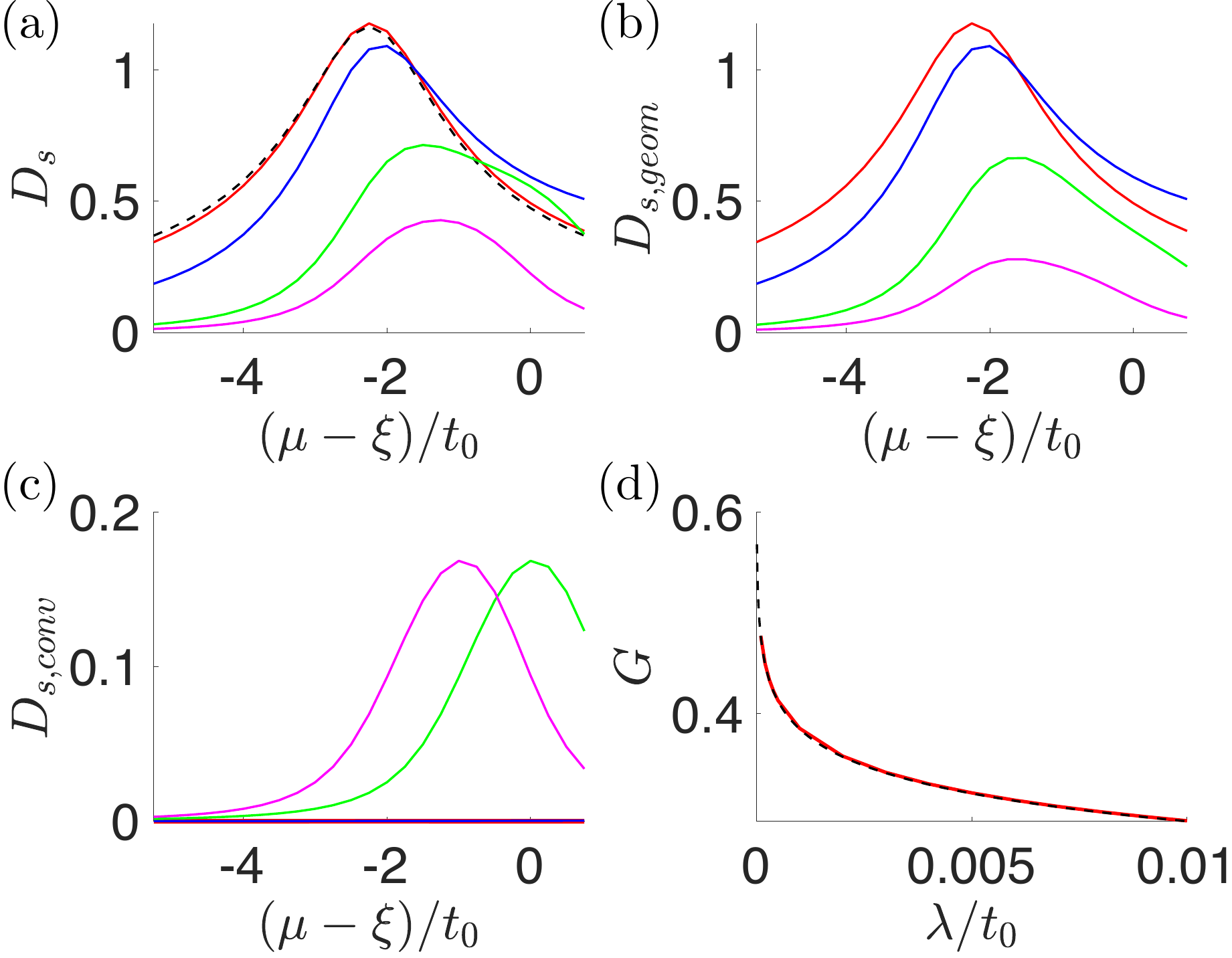}
\caption{(a) $D_s$ as a function of $\mu$ for the superconducting model (\ref{BdG-s-wave}) with $t_z=t_{z-}=0$, $\lambda=0.01 t_0$, $\Delta_0=t_0$, and other  hopping parameters given by the flat-band parameters in Table \ref{table:tb_params}. Additionally, we have introduced the perturbation (\ref{perturbation}), which shifts the lower band by an energy $\xi=0$ (magenta), $\xi=-t_0$ (green), $\xi=-10 t_0$ (blue) and $\xi=-100t_0$ (red)   without modifying the eigenstates. The dashed black line shows the analytic results (\ref{flat-band-geom-ideal}).  (b),(c) Same for $D_{s, geom}$ and $D_{s,conv}$, respectively. $D_{s,conv} \approx 0$ for all values of $\mu$ when $\xi/t_0 =: -100, -10$. (d) Integral of the quantum metric $G$ [Eq.~(\ref{G-def})] as a function of $\lambda$ (red line). Dashed black line shows the asymptotic behavior of $G$ [Eq.~(\ref{G-asymp})] when approaching the degeneracy of the bands ($\lambda \to 0$).}
    \label{fig:singlet-superfluidity-xi}
\end{figure}

In the case of time-reversal invariant $s$-wave singlet superconducting phase (\ref{BdG-s-wave}) the system obeys the spin-rotation symmetry around $z$-axis. Therefore, the corrections to the result (\ref{flat-band-geom-ideal}) only come from the facts that the bands are not well-isolated and they are not perfectly flat. We find that in this case the superfluid weight is large for all reasonable values of $\mu$ and $\Delta_0$ [Fig.~\ref{fig:singlet-superfluidity}(a)] and it contains a significant geometric component in the case of strong coupling superconductivity $\Delta_0 > 0.7 t_0$  [Fig.~\ref{fig:singlet-superfluidity}(b)]. The increase of the bandwidth caused by the H-M parameters relative to the flat-band parameters [see Fig.~\ref{fig:normal-state}(b)] increases the superfluid weight [see Fig.~\ref{fig:singlet-superfluidity}(c)] because the  conventional contribution is increased.  
The mirror symmetry breaking hopping ($t_z=t_0/12$) and spin-orbit coupling ($\lambda=0.3 t_0$) do not significantly affect the results for $D_s$ as shown in Fig.~\ref{fig:singlet-superfluidity}(c).

In the case of singlet superconductivity $D_s$ is fully determined by the quantum geometry (\ref{flat-band-geom-ideal}) in the limit where we have an isolated ($\xi=-100 t_0$) perfectly flat band ($t_z=0$) obtained with the help of the perturbation (\ref{perturbation}) as shown in Fig.~\ref{fig:singlet-superfluidity-xi}(a). When $\xi$ is decreased so that the bands are coupled the total superfluid weight $D_s$ slightly decreases in magnitude. The decrease of the geometric contribution $D_{s, geom}$ [Fig.~\ref{fig:singlet-superfluidity-xi}(b)] is partially compensated by the increase of conventional contribution $D_{s, conv}$ [Fig.~\ref{fig:singlet-superfluidity-xi}(c)]. We point out that the integral of the quantum metric
\begin{equation}
G=\frac{1}{(2\pi)^3} \int d^3k \, g_{xx}(\mathbf{k}) \label{G-def}
\end{equation}
can be very large in the our model. In fact, it diverges logarithmically as $\lambda \to 0$ [see Fig.~\ref{fig:singlet-superfluidity-xi}(d)]
\begin{equation}
G \approx 0.11+ \frac{1}{8 \pi} \ln(1/\lambda), \label{G-asymp}
\end{equation}
because the energy bands become degenerate along the $\Gamma$-$A$ line.
This divergence of $G$ does not, however, show up in the superfluid weight $D_s$ because the coupling of the bands becomes more and more important when approaching the degeneracy. 

\section{Conclusions}

We have computed the geometric and conventional contributions of the superfluid weight for the superconducting flat-band models proposed for copper-doped lead apatite. We have found that, in contrast to the standard paradigms of flat-band superconductivity, there does not exist any lower bound for the superfluid weight in these models. Moreover, although the nontrivial quantum geometries of the normal state bands are the same in the ferromagnetic and paramagnetic phases, the emerging superconducting phases have very different superfluid weights. Namely, we find that the ferromagnetic phase found in the DFT calculations could in principle support a variety of topologically nontrivial spin-polarized superconducting phases, but our results for the superfluid weight show that the their capabilities to support  supercurrents can vary a lot depending on the values of the model parameters. In particular,  there can even exist parameter regions where the homogeneous superconducting state becomes unstable. 
On the other hand, if the transition to the superconducting state is accompanied by the disappearance of spin polarization, the superfluid weight in the time-reversal invariant singlet superconducting state is robustly large and it contains a significant quantum geometric component. We have also shown that the spin-orbit coupling and mirror symmetry breaking terms in the Hamiltonian do not significantly affect the results for the superfluid weight even though they influence the normal state topology of the flat bands.

\section{Acknowledgements}
W.B.  acknowledges support by Narodowe Centrum Nauki 
(NCN, National Science Centre, Poland) Project No. 2019/34/E/ST3/00404 and the Foundation for Polish Science through the IRA Programme
co-financed by EU within SG OP.

\bibliography{bibliography}

\end{document}